\RequirePackage{fix-cm}
\documentclass[smallextended]{svjour3}       % onecolumn (second format)
\smartqed  % flush right qed marks, e.g. at end of proof
\usepackage{graphicx}
\usepackage{dsfont}
\usepackage{mathrsfs}
\usepackage{amssymb}
\usepackage{enumerate}
\usepackage{amsmath}
\usepackage{caption}
\usepackage{subfig}
\usepackage{appendix}
\usepackage[dvipsnames]{xcolor}
\usepackage{url}

\begin{document}

\title{Public Health Policy: COVID-19 Epidemic and SEIR Model with Asymptomatic Viral Carriers
}


\author{Weijie Pang       % \and
         %Stephan Sturm  %etc.
}


\institute{W. Pang \at
             Department of Mathematics and Statistics, McMaster University\\
%              Tel.: +123-45-678910\\
%              Fax: +123-45-678910\\
              \email{pangw6@mcmaster.ca}           %  \\
 %          \and
 %          S. Sturm \at
%              Department of Mathematical Science, Worcester Polytechnic Institute\\
  %            \email{ssturm@wpi.edu}
}

%\date{\today}%Received: date / Accepted: date}

\maketitle

\begin{abstract}
We measure the effect of different public health regulations to the spread of COVID-19, based on a SEIRA model -- a SEIR model including asymptomatic transmissions. The cumulative confirmed cases and death show nonlinear positive relationship with the value of asymptomatic rate. Based on this model, we analyze the inhibit effects to COVID-19 of three types of public health policies, i.e. isolation of laboratory confirmed cases, general personal protection and quarantine (lock-down). The simulations conclude that the isolation display limited effects to the asymptomatic viral carriers. The general personal protection and quarantine perform similar effects when the their percentages of participants are same. When the total proportion of asymptomatic, mild symptomatic and neglected patients is $40\%$, only depends on isolation policy may lead to an additional $75\%$ infections, compared with general personal protection or quarantine with an efficiency $80\%$. At end, we provide seven recommendations of public health intervention before and during an aerial transmitted epidemic (COVID-19).


\end{abstract}

\section{Introduction}

%{\color{red} COVID-19 prcess }COVID-19 start, situation, asymptomatic rate

Since the COVID-19 appeared in the late December, 2019 at Wuhan, China, it already infected more than 1,900,000 people in more than 185 countries. In Italy, there was two confirmed cases at January 31st. At same day, the government suspended all flights from China and  announced a national emergency. However, these actions did not stop the spread of the virus. 

Contrast with other Coronavirus, i.e. SARS and MERS, one essential different feature of COVID-19 is the asymptomatic infection. At the beginning of the spread of this virus, there were already several experts point out this. 
Many researchers \cite{al2020asymptomatic,bai2020presumed,chan2020familial,hu2020clinical,lai2020asymptomatic,tang2020detection,wang2020clinical} prove the existence of asymptomatic COVID-19 carriers by comparing the nucleic acid testing of patients' pharyngeal swab samples and chest CT scans with their typical symptoms and . Li et al. \cite{li2020early} mention that more young carriers show mild symptoms than older patients. Compared with retired people, young adult have more social activities in general. Include the asymptomatic carriers into an epidemic model is crucial. About the proportion of asymptomatic patients among all viral carrier, many experts \cite{wu2020characteristics,mizumoto2020estimating}  provide their own estimation based on distinct data source. 

For classical contagious disease, many models can predict the number of infections, recoveries and deaths. One important type is the compartmental model. Kermack and McKendrick \cite{kermack1991contributions} construct a deterministic model, called as SIR model, to quantify the susceptible population, infections and recoveries of one epidemic. Then they also updated this model in \cite{kermack1932contributions,kermack1933contributions}. Bj{\o}rnstad et al. \cite{bjornstad2002dynamics} apply this model to analyze the spread of Measles. McCluskey \cite{mccluskey2010complete} introduce the distributed delay and discrete delay into the SIR model. Some people update the model to a stochastic area \cite{ji2014threshold,jiang2011asymptotic,zhang2013stochastic}.
To include exposed people with latency period, Dietz \cite{dietz1975transmission} construct SEIR model. Based on SEIR model, many experts update this model \cite{anderson1979population,may1979population}
Many researchers use clinical and simulation data to calibrate the parameters in the epidemic model \cite{anderson1979population}. Various experts apply this type of model to analyze different diseases and develop it \cite{dukic2012tracking,lekone2006statistical,ozalp2011fractional}.  Ndii and Supriatna\cite{ndii2017stochastic} claim that a deterministic model is more appropriate than stochastic models for an large initial susceptible population.
More details of compartmental models in \cite{bailey1975mathematical,hethcote2000mathematics}.

% \cite{colizza2007modeling} use a stochastic epidemic model to study the spatial development of the H5N1 avian influenza virus. They find the relationship between air transportation with the global outbreak.

Several containment are proposed by experts.
Fraser et al. \cite{fraser2004factors} recommend to isolate symptomatic patients and their contacts to control an outbreak.  \cite{ferguson2006strategies} apply a large scale simulation to analyze the effect of publich health policy to the spread of one virus at the beginning period.  Wu et al. \cite{wu2006reducing} provide recommendation of household-based public health interventions.
Allred et al. \cite{allred2020regional} include the effect of asymptomatic patients to the demand of the health care. Bousema et al. \cite{bousema2014asymptomatic} discuss the detect tools of the asymptomatic Malaria infections with related public health. However, few paper focus on public health policy to inhibit an epidemic of aerial contagious virus with asymptomatic patients based on epidemic models.

The structure of this paper is as follow. Section \ref{sec:asssym} lists the symbols we need to construct our SEIRA model and the assumptions of to include the special characteristics of COVID-19. In Section \ref{sec:model}, we construct the SEIRA model and several related models to include the effects of different public health policy. In Section \ref{sec:simu}, we analyze the effects of asymptomatic rate, three types of public health containment (Confirmed Cases Isolation, Personal Protections, Quarantine), and delayed policy to the spread of COVID-19. Section \ref{sec:diss}, we provide the suggestions of public health agency for a new aerial transmitted virus.

\section{Assumptions \& Symbols}\label{sec:asssym}
In this paper, we have the following assumptions about COVID-19.
\begin{itemize}
\item No vaccine exist for COVID-19
\item People who recovered after catch COVID-19 will be insusceptible of it, all other people are susceptible. 
\item The number of infectious people only include the cases with typical symptoms and get confirmation from laboratory tests
\item The natural birth rate and natural death rate are same (the population is a constant)
\item The infectious ability of symptomatic and asymptomatic patients are same.
\end{itemize}

Before we introduce our time series SEIRA model, we list necessary symbols at first.
\begin{itemize}
\item $S$: Susceptible population
\item $E$: Population who are exposed to the virus and do get confirmed test
\item $I$: Population who get laboratory positive confirmation
\item $R$: Recovery (insusceptible to COVID-19)
\item $D$: Death number
\item $\alpha^I$: rate of people exposed to the infectious patients.
\item $\alpha^E$: rate of people exposed to the asymptomatic patients. 
\item $\beta$:  rate of exposed people catch COVID-19
\item $\beta^I$: rate of exposed people get COVID-19 with diagnosed confirmation
\item $\beta^E$: rate of exposed people without symptoms but can transmit it to others
\item $\gamma$:  recover rate of infected people 
\item $\mu^I$: death rate of symptomatic patients
\item $\mu^E$: death rate of asymptomatic patients
%\item $P_t$: the alternating renewal process of lock-down policy. $P_t = 1$ when the regulators start this policy
%\item $F_t$: the alternating renewal process of wearing face mask policy. $F_t = 1$ when the regulators start this policy
\item $e^Q$: Efficiency of quarantine (lock-down) policy
\item $e^F$: Efficiency of personal protection 
\item $e^L$: Efficiency of Isolation for confirmed cases
%\item $f(E_t)$: the function to describe the percentage of non-symptom patiences, assume it's a linear function so far $f(E_t) = \theta E_t, \theta < \alpha$, where $\theta$ is a constant rate of non-symptom patience with respect to the $E_t$
\end{itemize}

\section{Models}\label{sec:model}

\subsection{SEIR Model with asymptomatic Patients (SEIRA Model)}\label{sec:SEIRA}
Based on the classical SEIR model, we construct our SEIRA model. In this paper, we include all mild symptomatic and pre-symptomatic patients with capacity to transmit the virus in the number of asymptomatic viral carriers. The proportion of asymptomatic patients with respect to contagious population is 
\begin{equation}
r = \frac{\beta - \beta^I}{\beta}
\end{equation}
In common sense, testing broadly may help figure out those mild COVID-19 patient. But a special characteristic of COVID-19 is that there exists some asymptomatic patients. Therefore, to distinguish with other contagious model, we construct a model to include these sources of contagion.

%In the following model we apply the red term to include the non-symptom COVID-19 patience who still have capacity to transmise the virus to other people. 
\begin{equation}\label{md:SEIRA}
\begin{split}
\frac{dS_t}{dt} &= -\alpha^I S_t I_t  -\alpha^E \beta^E S_t E_t \\
\frac{dE_t}{dt} &= \alpha^I S_t I_t +\alpha^E \beta^E S_t E_t - \beta^I E_t -\mu^E \beta^E E_t\\
\frac{dI_t}{dt} &= \beta^I E_t - \gamma I_t -\mu^I I_t\\
\frac{dR_t}{dt} &= \gamma I_t \\
\frac{dD_t}{dt} &=\mu^I I_t + \mu^E \beta^E E_t\\
\end{split}
\end{equation}

\subsection{SEIRA Model with Public Health Policy}\label{sec:SEIRAP}
One important public health policy is self-isolating of all confirmed cases, which is also the first recommendation to their citizens for most government. However, considering the asymptomatic COVID-19 infectious persons, the self-isolating rule only control the movement of population $I$. As a result, this rule only decrease the $S_t I_t$ term in our SEIRA model. With a notation of $e^L$ as the rate of people following one policy, i.e. Isolation, our model is rewritten as
\begin{equation}\label{md:SEIRAP1}
\begin{split}
\frac{dS_t}{dt} &= -\alpha^I(1-e^L) S_t I_t  -\alpha^E \beta^E S_t E_t \\
\frac{dE_t}{dt} &= \alpha^I(1-e^L) S_t I_t +\alpha^E \beta^E S_t E_t - \beta^I E_t -\mu^E \beta^E E_t\\
\frac{dI_t}{dt} &= \beta^I E_t - \gamma I_t -\mu^I I_t\\
\frac{dR_t}{dt} &= \gamma I_t \\
\frac{dD_t}{dt} &=\mu^I I_t + \mu^E \beta^E E_t\\
\end{split}
\end{equation}

Another public health policy is general personal protection, including washing hands, general wearing mask, social distancing, which is an effective method to prevent both asymptomatic and symptomatic patients to transmit this virus. In other words, this policy may inhibit the rate of susceptible people's exposure to this virus, $\alpha^I$ and $\alpha^E$. The last and most strict public health regulation is the province-wide or nationwide quarantine (lock down). When we execute all three types, our model is
%In this paper, we denote the efficiency of wearing face mask as $e^F$, so the inhibitory effect of combination of self-isolating and wearing face mask to the spread of COVID is
\iffalse
\begin{equation}
\begin{split}
\frac{dS_t}{dt} &= -\alpha^I (1-e^F-e^L + e^Fe^L) S_t I_t  -\alpha^E (1-e^F)\beta^E S_t E_t \\
\frac{dE_t}{dt} &= \alpha^I (1-e^F-e^L + e^Fe^L) S_t I_t +\alpha^E(1-e^F) \beta^E S_t E_t - \beta^I E_t -\mu^E \beta^E E_t\\
\frac{dI_t}{dt} &= \beta^I E_t - \gamma I_t -\mu^I I_t\\
\frac{dR_t}{dt} &= \gamma I_t \\
\frac{dD_t}{dt} &=\mu^I I_t + \mu^E \beta^E E_t\\
\end{split}
\end{equation}
\fi
\begin{equation}\label{SEIRAP3}
\begin{split}
\frac{dS_t}{dt} &= -\alpha^I (1-e^F-e^L-e^Q + e^Fe^L+e^Fe^Q + e^Le^Q -e^Fe^Le^Q) S_t I_t  \\
&\qquad -\alpha^E (1-e^F-e^Q + e^Fe^Q)\beta^E S_t E_t \\
\frac{dE_t}{dt} &= \alpha^I (1-e^F-e^L-e^Q + e^Fe^L+e^Fe^Q + e^Le^Q -e^Fe^Le^Q) S_t I_t \\
&\qquad +\alpha^E(1-e^F-e^Q + e^Fe^Q) \beta^E S_t E_t - \beta^I E_t -\mu^E \beta^E E_t\\
\frac{dI_t}{dt} &= \beta^I E_t - \gamma I_t -\mu^I I_t\\
\frac{dR_t}{dt} &= \gamma I_t \\
\frac{dD_t}{dt} &=\mu^I I_t + \mu^E \beta^E E_t\\
\end{split}
\end{equation}

\section{Simulations}\label{sec:simu}

Because we want to check the inhibitory effects to the spread of COVID-19 for different efficiencies of public health policy, all rate $e^L$, $e^F$, and $e^Q$ would run between $0\%$ to $95\%$  In our simulation, we cite the data from John Hopkins Coronavirus Resource Center (JHU CSSE) \cite{jhdata} as source. Based on the data at 4:00 pm on April 5$^{th}$, 2020, the case fatality rate (CFR) for countries with Top 6 most confirmed cases are $12.32\%$ (Italy), $9.5\%$ (Spain), $8.34\%$ (France), $4.04\%$ (China) $2.85\%$ (U.S.), $1.54\%$ (Germany). In our model, we assign the constant death rate $\mu^I$ equal the average $6.43\%$ and $\mu^E$ equal 0, which means all people died of COVID-19 should be laboratory-confirmed positive. 

Del Valle et al. \cite{del2007mixing} conclude that the contact numbers for population are distinct from different age group, arriving peak (about 21 average number of contacts per person daily) between 20 and 50 years old. In this paper, we take a rough mean as 15. So we assume the exposure rate $\alpha^I = \alpha^E = 15/S_0$, where $S_0$ is the initial susceptible population.

The reproduction number ($R_0$) is the average number of people catch this disease, caused by one infectious people. Liu et al. \cite{liu2020reproductive} state the mean reproduction number ($R_0$) of COVID-19 is around 3.28, with a median of 2.79 and a interquartile range (IQR) of 1.16.  Zhang et al. \cite{zhang2020estimation} claim the $R_0$ is 2.28 with a $95\%$ confident interval as $(2.06, 2.52)$. In this paper, we use the average $R_0 = 2.54$. Follow the findings from Li et al. \cite{li2020early}, Lauer et al. \cite{lauer2020incubation} and Anderson et al \cite{anderson2020will}. In this model, we denote the incubation period as $d$ days, with assuming that the average incubation period of COVID-19 is 5.5 days. With the value of $R_0$ and relationship among $R_0$, $\beta_I$ and the average incubation period by \cite{colizza2007modeling,diekmann2000mathematical}, the estimation of $\beta^I$ is 
\begin{equation}
\beta^I = \frac{R_0}{\alpha^I \times d} = 3.08\%
\end{equation}

%, the percentage of asymptomatic patients in all COVID-19 patients is between $19\%$ and $31\%$. If one government does not apply general testing policy, the COVID-19 patient without positive test result can even higher. As a result, we set the asymptomatic rate $r$ run from $10\%$ to $60\%$. So our parameter $\beta^E = \frac{\alpha^I r}{1-r} *(1-\gamma)$, where we assume $\gamma = 0.5$.

\subsection{Effect of asymptomatic Rate}\label{sec:asymp}

In this section, we want to simulate the spread of COVID-19 with respect to the asymptomatic rate $r$. To avoid the effect of any public health policy and any other viral source, we assume there is no input confirmed case and no inhibit policy. 
We are interested in the effect of $r$ to the epidemic is that, COVID-19 show much higher percentage of asymptomatic and mild symptomatic patients. In our model, the infectious number $I$ is only include the laboratory confirmed cases. Because the asymptomatic patients and people in incubation period can also infect susceptible people, we want to check the effect of the asymptomatic rate $r$ to the spread of COVID-19. The research in this topic may explain the distinct performance of COVID-19 with other virus, i.e. SARS, MERS, and N1H1.

Many experts provide their estimation of the asymptomatic rate $r$. Mizumoto et al. \cite{mizumoto2020estimating} studied the 3,063 tests on board the Diamond Princess cruise ship at Yokohama, Japan and got that the estimation of the asymptomatic proportion was $17.9\%$. Wu and McGoogan \cite{wu2020characteristics} the rate is only $1\%$. In face, finding a viral carrier without symptoms is really hard. Although, the clinical tools, such as testing the nucleic acid of the pharyngeal swab sample and chest CT scans, can help doctors, but the finding really depends on a local test capability, a personal opinion of infectious people and the incubation period of a virus. In general, when some patients ignore their mild symptoms at the beginning or a potential carrier cannot get tested immediately, the value from existing data is much underestimated. So in our paper, we analysis the effect of the spread of COVID-19 for the asymptomatic rate $r$ between $0\%$ and $60\%$.

\begin{figure}[h!]
  \subfloat[Cumulative Confirmed Case]{
	\begin{minipage}[c][1\width]{
	   0.5\textwidth}
	   \centering
	   \includegraphics[width=1\textwidth]{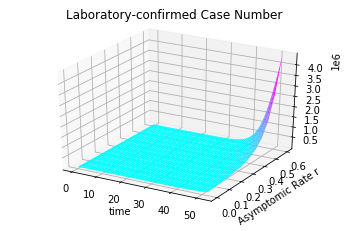}
	\end{minipage}}
 \hfill 	
  \subfloat[Total Death]{
	\begin{minipage}[c][1\width]{
	   0.5\textwidth}
	   \centering
	   \includegraphics[width=1.1\textwidth]{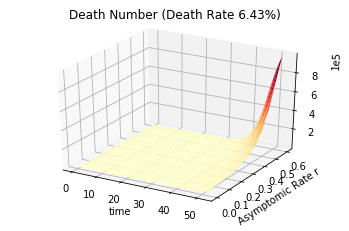}
	\end{minipage}}
\caption{Spread of COVID-19 Within 52 Days}
\label{fig:rd52}
\end{figure}

As early as January 31, Italy declared a national emergency after two confirmed cases in Rome. However, except a ban of flights to China, Italy didn't take any other classical public health policy. Until March 9th, Italy locked down the whole country, which is 38 days after the first two confirmed case. Assuming there was only one viral carrier on January 31, we want to check the spread of COVID-19 without any inhibit regulation after additional 14 days. Figure \ref{fig:rd52} show the increasing number  of cumulative confirmed case and death for asymptomatic rate $r\leq 60\%$ within 52 days. From this figure, we can conclude that the asymptomatic viral carriers can aggravate the transmission dramatically.

%\begin{figure}[h!]
%\centering
%\includegraphics[width = 1\textwidth]{graph/Italy_confirm_asymptom_oned52}
%\caption{Confrimed Case Number After 52 Days}
%\label{fig:roned52}
%\end{figure}

\begin{figure}[h!]
  \subfloat[Cumulative Confirmed Case]{
	\begin{minipage}[c][1\width]{
	   0.5\textwidth}
	   \centering
	   \includegraphics[width=1\textwidth]{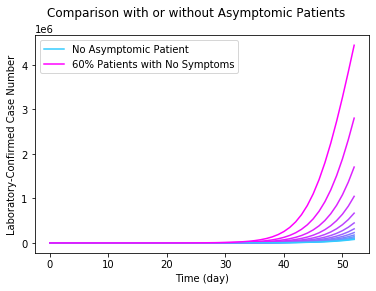}
	\end{minipage}}
 \hfill 	
  \subfloat[Total Death]{
	\begin{minipage}[c][1\width]{
	   0.5\textwidth}
	   \centering
	   \includegraphics[width=1.1\textwidth]{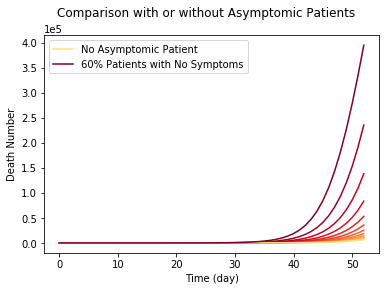}
	\end{minipage}}
\caption{Comparison for Asymptomatic Rate $r = 0\%$ and $r = 40\%$}
\label{fig:roned}
\end{figure}

In Figure \ref{fig:roned}, we focus one the two case, $r = 0\%$ and $r = 40\%$.  When $r=0$, the cumulative confirmed case number would be 3592 at March 9th (38 days after the time of first confirmed case). This number is much lower than the Italy published case number 9172. But for $r=40\%$, the cumulative confirmed case number is 17441, which is nearly double the published number. Therefore, it's important to include the transmission from asymptomatic viral carriers into the epidemic model. Then, the death rate is 294 when $r = 0\%$, 1348 when $r = 40\%$, and the public number 463. In our model, we assume all patients would have typical symptoms before they die, which is contradict to the real life. Since the testing rate in Italy is not ideally at the beginning, some patient didn't have opportunity to receive laboratory result before everything are too late. As a result, it is reasonable that the simulated total death is three times the published number.

When Italy government realize the severe condition of COVID-19, they closed schools and universities on March 4th, locked down some provinces at northern part and then extended the lock-down policy to the whole country on March 9th. However, if the government didn't take any action, the cumulative confirmed case number would reach $669,115$ on March 23th when $r = 40\%$, just two additional weeks. This is 38 times the number with same parameters on March 9th. When $r = 0\%$, the simulated result is 78,306, which is not far away from the published number 63,927. 
Unfortunately, medical case study already proved that the asymptomatic rate is not zero for COVID-19. Considering viral carriers can transmit it during the incubation period, the published number is much lower than the real existing cases in the country. The lack of testing at the beginning of this epidemic, the regulator extremely underestimate the severe level.

\subsection{Effect of Single Public Health Policy}\label{sec:policy}

In this section, we discuss the inhibition effects of different public health policies. Here we focus on three rules,  the self-isolated rule of all laboratory confirmed patients, wearing face masks (covering noses and mouths) for all people, and nationwide lock-down (quarantine rule).

Some experts believe that ignoring asymptomatic feature have minor change to the prediction of a epidemic \cite{colizza2007modeling}. Statistically, we can have similar result, if we use a larger estimation of the mean reproduction number $R_0$. Unfortunately, this neglect lead to different regulation. Without asymptomatic carriers, the transmission of virus can be stopped by isolating all active confirmed patients. But this rule cannot forbid the asymptomatic or mild symptomatic patients to transmit the virus. As a result, we analyze the effect of these regulations with different setting of the asymptomatic rate $r$. 

\subsubsection{Effect of Self-Isolation}\label{sec:eL}

\begin{figure}[h!]
  \subfloat[Immediate Isolation when $r = 0\%$]{
	\begin{minipage}[c][1\width]{
	   0.5\textwidth}
	   \centering
	   \includegraphics[width=1\textwidth]{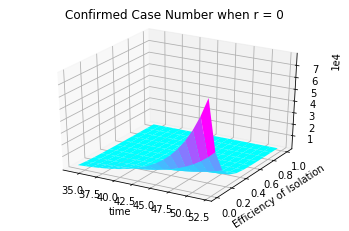}
	\end{minipage}}
  \subfloat[Immediate Isolation when $r = 40\%$]{
	\begin{minipage}[c][1\width]{
	   0.5\textwidth}
	   \centering
	   \includegraphics[width=1\textwidth]{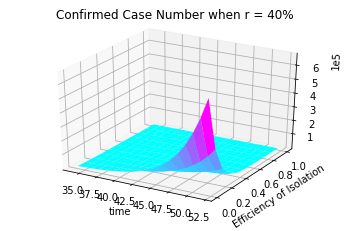}
	\end{minipage}}
\caption{Effect of Self-Isolation to Cumulative Confirmed Cases}
\label{fig:isoconfirm}
\end{figure}

In this part, we discuss the inhibition of the epidemic by the self-isolation rule of laboratory confirmed cases. Here we have two start days of the self-isolation policy, starting at the initial day (January 31st) or starting on March 9th. Figure \ref{fig:isoconfirm} shows the increase of the cumulative confirmed cases between March 5th (34th day) and March 23rd (52rd day). %and Figure \ref{fig:isodeath} shows the change of total death during same period.

%This is also important for the analysis of public health policy. In the beginning of the epidemic, most government only require the laboratory confirmed patient follow the self-isolation rule. In other word, the viral carrier without testings or symptoms are allowed to participate in normal social activities. 

\begin{figure}[h!]
  \subfloat[Delayed Isolation when $r = 0\%$]{
	\begin{minipage}[c][1\width]{
	   0.5\textwidth}
	   \centering
	   \includegraphics[width=1\textwidth]{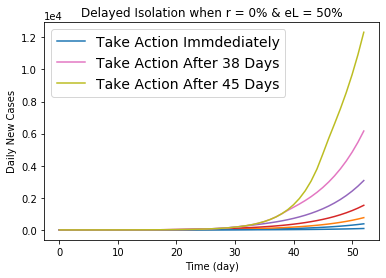}
	\end{minipage}}
  \subfloat[Delayed Isolation when $r = 40\%$]{
	\begin{minipage}[c][1\width]{
	   0.5\textwidth}
	   \centering
	   \includegraphics[width=1\textwidth]{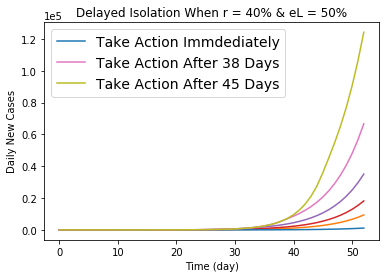}
	\end{minipage}}
\caption{Effect of Delayed Isolation to Daily New Cases ($e^L = 50\%$)}
\label{fig:isodelay}
\end{figure}

By these two figures, the self-isolation policy can inhibit the spread of COVID-19, and the effect show positive nonlinear relationship with the proportion of the population.  Figure \ref{fig:isoconfirm} a) and b) show the effect of different self-isolate levels, which are applied immediately on January 31st. Contrarily, the self-isolation rule is delayed by 10, 17, 24, 31, 38, or 45 days in Figure \ref{fig:isodelay} a) and b). For any estimated asymptomatic rate $r$ and execute date, the Self-isolation rule inhibit the spread of COVID-19. 

But when we zoom in the picture, we find the efficiency of the rule is better for a low asymptomatic rate and an early enforce date. Figure \ref{fig:isodelay} show the difference between a self-isolation rule start on January 31 (Day 0) and delayed time to the daily new cases. For the fast action, only $50\%$ laboratory-confirmed case avoid all social contacts and self-isolate themselves, which help the whole country avoid the epidemic from the beginning. However, if the isolation rules was execute fully at Day 38 (Light Pink Curve), the daily new case would be 1,210 on March 23rd (Day 52) with $e^L = 100\%$ and 6,162 on Day 52 for $e^L = 50\%$ with $r = 0\%$. The situation is even worse for COVID-19 with asymptomatic patients, there would be 14,972 people catch it on March 23rd with the delayed reaction at Day 38 from public health regulators.  In general, every week delay of taking actions, involving half objects, may lead to double cumulative confirmed cases in a period of 52 days.

\subsubsection{Effect of General Personal Protection}\label{sec:eF}

\begin{figure}[h!]
  \subfloat[Immediate Policy when $r = 0\%$]{
	\begin{minipage}[c][1\width]{
	   0.5\textwidth}
	   \centering
	   \includegraphics[width=1\textwidth]{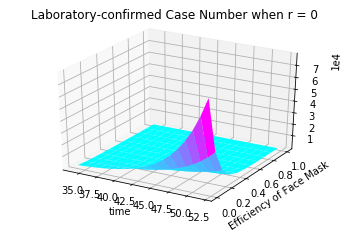}
	\end{minipage}}
  \subfloat[Immediate Policy when $r = 40\%$]{
	\begin{minipage}[c][1\width]{
	   0.5\textwidth}
	   \centering
	   \includegraphics[width=1\textwidth]{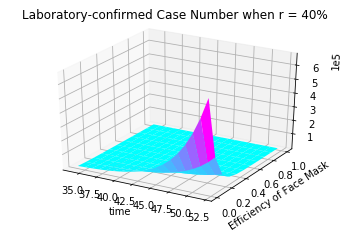}
	\end{minipage}}
\caption{Effect of General Personal Protections to Cumulative Confirmed Cases}
\label{fig:fmconfirm}
\end{figure}

Now, we measure the inhibition of the epidemic by requiring all people take action to protect themselves. Similarly with the discussion of self-isolation policy, we also have different start days of this rule. Figure \ref{fig:fmconfirm} shows the increase of the cumulative confirmed cases between March 5th (34th day) and March 23rd (52nd day), which prove the inhibition effect of general personal protections.

\begin{figure}[h!]
  \subfloat[Delayed Policy when $r = 0\%$]{
	\begin{minipage}[c][1\width]{
	   0.5\textwidth}
	   \centering
	   \includegraphics[width=1\textwidth]{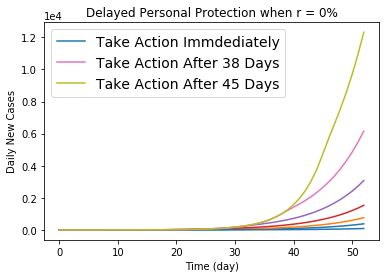}
	\end{minipage}}
  \subfloat[Delayed Policy when $r = 40\%$]{
	\begin{minipage}[c][1\width]{
	   0.5\textwidth}
	   \centering
	   \includegraphics[width=1\textwidth]{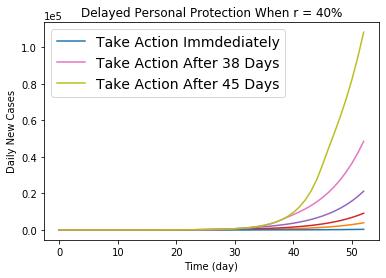}
	\end{minipage}}
\caption{Effect of Delayed General Personal Protections to Daily New Cases}
\label{fig:fmdelay}
\end{figure}

Comparing the effect of immediate execute policy in Figure \ref{fig:fmconfirm} a), b) with delayed starting time in Figure \ref{fig:fmdelay} a), b), taking action earlier also provide better effect to the control of the transmission for any asymptomatic rate $r$. Consistence with the self-isolation rule of diagnosed cases, a stricter public health policy brings a sooner controlled situation.

\iffalse
\begin{figure}[h!]
  \subfloat[Percentage storage utilization]{
	\begin{minipage}[c][1\width]{
	   0.5\textwidth}
	   \centering
	   \includegraphics[width=1\textwidth]{graph/Italy_death_eF_d3452_r0.png}
	\end{minipage}}
 \hfill 	
  \subfloat[Percentage storage utilization]{
	\begin{minipage}[c][1\width]{
	   0.5\textwidth}
	   \centering
	   \includegraphics[width=1\textwidth]{graph/Italy_death_eF_d3452_r0_sd38.png}
	\end{minipage}}
 \hfill 	
  \subfloat[standard deviation]{
	\begin{minipage}[c][1\width]{
	   0.5\textwidth}
	   \centering
	   \includegraphics[width=1.1\textwidth]{graph/Italy_death_eF_d3452_r04.png}
	\end{minipage}}
  \subfloat[Percentage storage utilization]{
	\begin{minipage}[c][1\width]{
	   0.5\textwidth}
	   \centering
	   \includegraphics[width=1\textwidth]{graph/Italy_death_eF_d3452_r04_sd38.png}
	\end{minipage}}
 \hfill 	
\caption{Effect of Self-Isolation Policy for Difference asymptomatic Rates}
\end{figure}

\begin{figure}[ht]
  \subfloat[Percentage storage utilization]{
	\begin{minipage}[c][1\width]{
	   0.5\textwidth}
	   \centering
	   \includegraphics[width=1\textwidth]{graph/Compare_eF_asymptom}
	\end{minipage}}
 \hfill 	
  \subfloat[standard deviation]{
	\begin{minipage}[c][1\width]{
	   0.5\textwidth}
	   \centering
	   \includegraphics[width=1.1\textwidth]{graph/Compare_eF_asymptom_sd38}
	\end{minipage}}
\caption{}
\end{figure}
\fi

\subsubsection{Effect of Quarantine (Lock Down) Policy}\label{sec:eQ}

\begin{figure}[h!]
  \subfloat[Immediate Quarantine when $r = 0\%$]{
	\begin{minipage}[c][1\width]{
	   0.5\textwidth}
	   \centering
	   \includegraphics[width=1\textwidth]{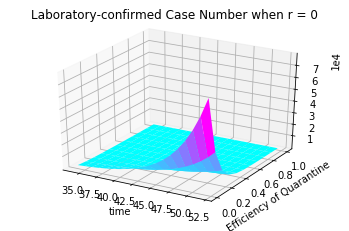}
	\end{minipage}}
  \subfloat[Immediate Quarantine when $r = 40\%$]{
	\begin{minipage}[c][1\width]{
	   0.5\textwidth}
	   \centering
	   \includegraphics[width=1\textwidth]{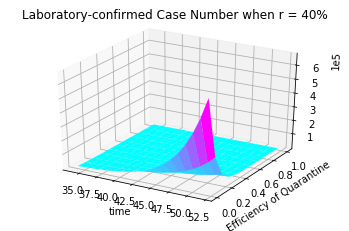}
	\end{minipage}}
\caption{Effect of Quarantine to Cumulative Confirmed Cases}
\label{fig:qconfirm}
\end{figure}

\begin{figure}[h!]
  \subfloat[Delayed Quarantine when $r = 0\%$]{
	\begin{minipage}[c][1\width]{
	   0.5\textwidth}
	   \centering
	   \includegraphics[width=1\textwidth]{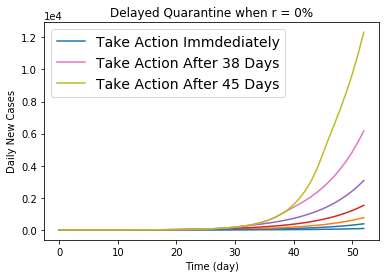}
	\end{minipage}}
 \subfloat[Delayed Quarantine when $r = 40\%$]{
	\begin{minipage}[c][1\width]{
	   0.5\textwidth}
	   \centering
	   \includegraphics[width=1\textwidth]{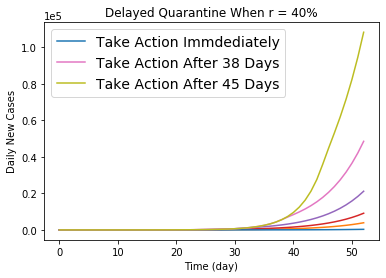}
	\end{minipage}}
\caption{Effect of Delayed Quarantine to Daily News Cases}
\label{fig:qdelay}
\end{figure}

In this part, we move to a discussion of the most tight public health intervention to stop the transmission of the virus. Figure \ref{fig:qconfirm} shows the effect with respect to different asymptomatic rates $r$ of a virus and executed levels. Figure \ref{fig:qdelay} explains later implement date of the quarantine to the daily increase of cases between. Similarly with the other two policies, this effects also has positive nonlinear relationships with earlier implement, stricter rules and lower asymptomatic rate.

\iffalse
\begin{figure}[h!]
  \subfloat[Percentage storage utilization]{
	\begin{minipage}[c][1\width]{
	   0.5\textwidth}
	   \centering
	   \includegraphics[width=1\textwidth]{graph/Italy_death_eQ_d3452_r0.png}
	\end{minipage}}
 \hfill 	
  \subfloat[Percentage storage utilization]{
	\begin{minipage}[c][1\width]{
	   0.5\textwidth}
	   \centering
	   \includegraphics[width=1\textwidth]{graph/Italy_death_eQ_d3452_r0_sd38.png}
	\end{minipage}}
 \hfill 	
  \subfloat[standard deviation]{
	\begin{minipage}[c][1\width]{
	   0.5\textwidth}
	   \centering
	   \includegraphics[width=1.1\textwidth]{graph/Italy_death_eQ_d3452_r4.png}
	\end{minipage}}
  \subfloat[Percentage storage utilization]{
	\begin{minipage}[c][1\width]{
	   0.5\textwidth}
	   \centering
	   \includegraphics[width=1\textwidth]{graph/Italy_death_eQ_d3452_r4_sd38.png}
	\end{minipage}}
 \hfill 	
\caption{Effect of Self-Isolation Policy for Difference Asymptomatic Rates}
\end{figure}

\begin{figure}[ht]
  \subfloat[Percentage storage utilization]{
	\begin{minipage}[c][1\width]{
	   0.5\textwidth}
	   \centering
	   \includegraphics[width=1\textwidth]{graph/Compare_eQ_asymptom}
	\end{minipage}}
 \hfill 	
  \subfloat[standard deviation]{
	\begin{minipage}[c][1\width]{
	   0.5\textwidth}
	   \centering
	   \includegraphics[width=1.1\textwidth]{graph/Compare_eQ_asymptom_sd38}
	\end{minipage}}
\caption{}
\end{figure}
\fi

\subsection{Comparisons of Different Policies}\label{sec:compare}
\begin{figure}[h!]
  \subfloat[Comparison when $r = 0\%$]{
	\begin{minipage}[c][1\width]{
	   0.5\textwidth}
	   \centering
	   \includegraphics[width=1\textwidth]{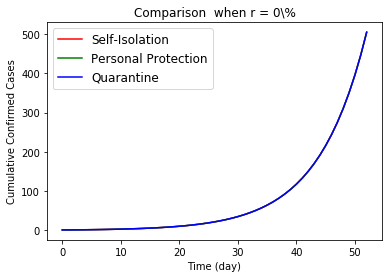}
	\end{minipage}}
 \hfill 	
  \subfloat[Comparison when $r = 40\%$]{
	\begin{minipage}[c][1\width]{
	   0.5\textwidth}
	   \centering
	   \includegraphics[width=1.1\textwidth]{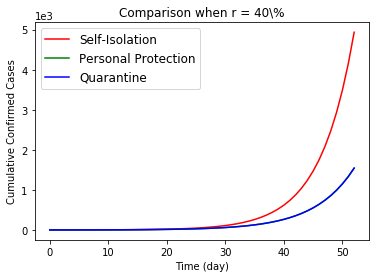}
	\end{minipage}}
\caption{Comparison of the Effects to Cumulative Confirmed Cases}
\label{fig:comparedays}
\end{figure}

The simulation already show the positive effect of these three public health policy. In this section, we want to compare the effect of these three policies. Figure \ref{fig:comparedays} draw the effects to the cumulative confirmed cases by three regulations. When a virus do not have asymptomatic carrier, there is not different among the polices. However, the personal protections and quarantine are better than self-isolation of confirmed patients when we consider the asymptomatic infections for COVID-19 epidemic. Reviewing the intense condition of all countries involved in the COVID-19 epidemic, it is not reasonable to suppose patients can receive tests and treatments on time during a general broadly viral transmission. So, the asymptomatic viral population contains all mild symptomatic patients and infectious patients who cannot received laboratory test on time. Therefore, we conclude only self-isolating the laboratory confirmed case is worse than the other two, when regulators try to control the spread of one aerial viral epidemic.

\begin{figure}[ht]
  \subfloat[When $r = 0\%$]{
	\begin{minipage}[c][1\width]{
	   0.5\textwidth}
	   \centering
	   \includegraphics[width=1\textwidth]{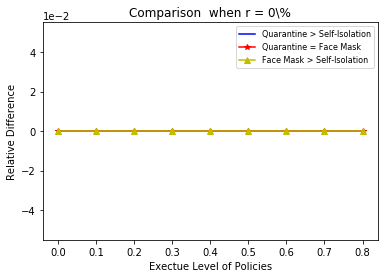}
	\end{minipage}}
 \hfill 	
  \subfloat[When $r = 40\%$]{
	\begin{minipage}[c][1\width]{
	   0.5\textwidth}
	   \centering
	   \includegraphics[width=1.1\textwidth]{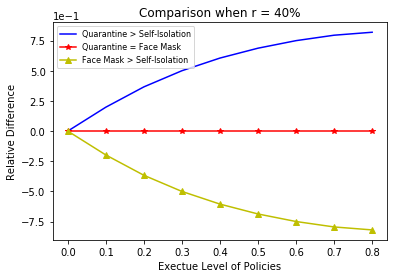}
	\end{minipage}}
\caption{Relative Different Effect to Daily New Cases}
\label{fig:comparenew}
\end{figure}

Then the general personal protection at beginning is better than a quarantine, although they show similar effects. For quarantine, even though all citizens support this policy, many people have to work and contact with others to maintain the function of our society, i.e. staff who are working at hospitals, groceries, post office, waste collection, public transportation, airports and so on. So in Figure \ref{fig:comparenew}, we only compare the relative difference effects to the daily new cases from no action to $90\%$ proportional execution. Worse, prohibition of all people for any social and daily activities would cause incredible interruption to the society. 

%\subsubsection{Effect of Combined Public Health Policy}

\section{Discussion}\label{sec:diss}

In previous sections, we already construct an epidemic SEIRA model and use the simulation analyze the effects of confirmed case isolation, personal protection and quarantine. In this section, we want to give a guideline about the non-pharmaceutical public health interventions before and during a pandemic.

In this paper, we assume that no imported viral source to the system. In general, a ban for other countries only delay the peak, instead of prevent it. Ferguson et al. \cite{ferguson2006strategies} find a 99.9\% ban can postpone 6 weeks for United States.  Therefore, we do not include the travel restrictions into our public health policy. We want to discuss the preparation for a government during this period.

First of all, at the start of one pandemic, the government must pay attention to the asymptomatic rate. Compared the cumulative confirmed cases and deaths with low and high asymptomatic rate in Section \ref{sec:asymp}, the proportion of asymptomatic viral carriers (with mild symptomatic patients) among all infections is an influential parameter of any viral transmission, also supported by Fraser Fraser et al. \cite{fraser2004factors}. When there is limited local casts, the researchers must pay attention to foreign researchers result about asymptomatic rate. If the pandemic start at our country, health workers need to estimate the rate as soon as possible. 

Next, more tests are better. In general, at the beginning of one transmission, only symptomatic patients get tests. However, doctors cannot find asymptomatic patients in this situation, until one asymptomatic patient show typical symptoms. This test rule lead to overlook of asymptomatic infections. In fact, we should tests all contacts of every confirmed cases. Based on the results of these test, we can estimate the reproduction number $R_0$ and the asymptomatic rate $r$. Moreover, doctors should have the right to test potential infections with typical symptoms even though they do not have direct contacts with confirmed cases or travel history. As we all known, a dangerous sign of pandemic is community transmission without clear source, contradict to the previous rule. National Public Health Agency should permit doctors to do test for more patients without direct contacts or travel history.

For laboratory confirmed cases, the compulsory self-isolation must apply to all direct contacts compulsory immediately. In Section \ref{sec:asymp}, we also analyzed the spread of one virus if we miss one diagnosed case initially. In Section \ref{sec:eL}, we discussed the sensitivity of the effects for the policy to the epidemic. Besides test all direct contacts, the isolation of the household immediately is also necessary. 

Facing a new virus, the government should recommend general public take personal protection as soon as possible. When no information about an asymptomatic rate $r$, experts should assume there is asymptomatic patients. Compared the death and cumulative confirmed cases in Section \ref{sec:asymp}, it is really dangerous to ignore this viral transmission. Because isolation of confirmed cases is hard to control it, a general personal protection is one of the best method. In addition, this method is also better than lock down policy. Although these two methods share similar inhibit effects, personal protection lead to fewer disruptions to the society and economic. Public health agency need to involve general public into the fight with epidemic as soon as possible, in order to avoid having to implement quarantine province-wide or nationwide.

For general personal protections, there are several types, such as social distancing, washing hands, wearing mask and others. In these three methods, the first two are much easier than the third one. Washing hand is the most useful and economical method to protect people. However, when the virus is transmitted by droplets or aerosols, only depends on washing hands could lead to dreadful result. In this case, social distancing is essential. Unfortunately, the social distance require critical condition about the location. In many place, such as the vehicle of most public transportation, school, and many workplace, no enough space exists to keep social distance. In this situation, general mask-wearing is a good substitution. Tellier \cite{tellier2006review} recommend use N95 respirators as part of personal protective equipment for aerosol transmission. Considering the storage amount and price of N95, it should be saved to people work in high risk area. Li et al. \cite{li2006vivo} the efficiency of surgical masks is 95\%, compared with 97\% for N95. As a result, an optimal choice for general public is surgical mask during an outbreak. Because of the paucity of research about homemade face masks, we cannot give any conclusion about homemade face mask in this paper. Before cumulative cases and death increase sharply, the government need take action to produce or purchase enough surgical and N95 face masks for both health workers and general population, and educate people about the correct method to wear a face mask and protect themselves on the same time. During a aerial transmitted disease, a combination of all three types of personal protections is indispensable.

Last but not least, government need lock down an outbreak zone determinedly. Although any lock-down policy brings horrible outcomes in economic and society, it is the final policy to save life. In section \ref{sec:compare}, we already compare isolation, personal pretection and quarantine. An effective personal protection should show similar consequence with a quarantine. But when a disease is out of control, it means the rule of isolation and general personal action are too late or too liberal. In this situation, the government has to execute more rigorous policy to inhibit the transmission, i.e. closing school, cancellation of public gatherings, working from home, limited mass transportation, and national quarantine.

\section{Conclusion}\label{sec:con}
In this paper, we construct two epidemic models. Firstly, the SEIRA model can be used to predict epidemic with asymptomatic viral carriers. Then to analyze the effect of different public health intervention, we use a SEIRA with policy model. By simulation, we find that asymptomatic rate $r$ is crucial parameters for the health policy. When $r \geq 10\%$, government need to include general personal protection into their public health policy, especially wearing-mask for aerial transmitted virus. When there is a new virus with transmission by droplets or aerosols, the government must 
\begin{itemize}
\item Test all typical symptomatic suspects and direct contacts of confirmed cases
\item Estimate the proportion of asymptomatic viral carriers
\item Prepare sufficient N95 and relative equipment for health worker
\item Prepare enough surgical face masks for general public
\item Execute compulsory isolation of confirmed case and all contacts
\item Educate people with correct personal protection methods at the beginning
\item Enforce quarantine when an epidemic is unavoidable
\end{itemize}

\bibliographystyle{spmpsci}
\bibliography{references_thesis}

\end{document}